\documentclass[twocolumn,aps,pra,a4paper,showpacs,nofootinbib,10pt]{revtex4-1}
\usepackage{graphicx}
\usepackage{keyval2e}
\usepackage[caption=false]{subfig}
\usepackage{amsmath}
\usepackage{bbold}
\usepackage{color}
\graphicspath{{CTMfigures/}}
\usepackage[ruled]{algorithm2e}

\newcommand{\ket}[1]{|#1\rangle}

\begin{document}

\title{Cross-talk minimising stabilizers} \author{Kaila
  C. S. Hall} \email{kaila.hall@strath.ac.uk} \affiliation{SUPA
  Department of Physics, University of Strathclyde, Glasgow, G4 0NG,
  United Kingdom}

\author{Daniel K. L. Oi} \affiliation{SUPA Department of Physics,
  University of Strathclyde, Glasgow, G4 0NG, United Kingdom}

\begin{abstract}
  Verification and characterisation of quantum states are crucial for
  the implementation of quantum information processing, especially for
  many-body systems such as cluster states in optical lattices.  In
  theory, it is simple to estimate the distance of a state with a
  target cluster state by measurement of a set of suitable stabilizer
  operators. However, experimental non-idealities can lead to
  complications, in particular cross-talk in single site addressing
  and measurement. By making a suitable choice of stabilizer operator
  sets we may be able to reduce, but not eliminate, these cross-talk
  errors. The degree of cross-talk mitigation depends on the geometry
  of the cluster state and subsets of cross-talk free stabilizers can
  be generated for certain shapes using a simple algorithm.
\end{abstract}

\date{\today}


\maketitle

\section{Introduction}

Cluster states, highly entangled many-body systems of qubits, are a
resource for quantum information processing through a sequence of
local measurements and
feedforward~\cite{PhysRevA.68.022312,PhysRevLett.86.5188} and their
generation is a highly active area of
research~\cite{PhysRevA.73.033818,PhysRevA.79.052324,OptCom.281.5282,NatCom.4.2161,Nature.425.937,Nature.434.169,PhysRevA.85.052305}. A
leading candidate for measurement-based quantum computation is an
array of atoms trapped by an optical lattice and recent experiments
have demonstrated individual qubit addressing~\cite{Nature.471.319}.

In addition to the experimental challenge of generating cluster
states, it is especially difficult to verify and characterise such a
many-body quantum
system~\cite{quantumstateestimation,quantph9807006v1} where quantum
tomography~\cite{PhysRevA.40.2847} is infeasible. More efficient ways
of determining key characteristics are required, for example
entanglement can be detected using an entanglement
witness~\cite{PhysLettsA.271.319}. It is also important to know how
close the actual state produced is to the desired one. For cluster
states, theoretically it is a simple matter to compare the expectation
values of the set of stabilizer operators that describe the desired
cluster state with the measured results~\cite{PhysRevA.68.022312,PhysRevLett.86.5188,PhysRevA.74.052302}.

Measuring such quantities may not be straightforward
however. Typically, the spacing of atoms in an optical lattice is of
the same order as the wavelength of light used to individually address
them which may lead to cross-talk~\cite{Nature.471.319}. Such
cross-talk can cause errors in the measured operators hence reducing
the accuracy of the distance estimation. Composite or compensation
pulse sequences, as carried out in NMR could be used to correct for
systematic errors, but it may be desirable to avoid the complexity and
overhead this introduces~\cite{PhysRevA.67.042308}.

Instead of using the standard (or canonical) set of stabilizers to
characterize a cluster state, we instead seek to find alternative
descriptions that can reduce the problem of cross-talk. We achieve
this by finding some stabilizers that eliminate the need to perform
local addressing as well as others that minimize the number of
neighbouring measurements in different bases. We find such cross-talk
minimizing sets for a range of different cluster state geometries, in
particular square and triangular lattices.

\subsection{Cluster States}

A cluster state is a many body quantum system defined as the
simultaneous $+1$ eigenstate of a set $\mathcal{S}$ of commuting
stabilizer operators $\hat{S}^a$~\cite{PhysRevA.68.022312},
\begin{equation}
\hat{S}^a\ket{\psi}=+\ket{\psi},\;\forall\hat{S}^a\in\mathcal{S}.
\end{equation}
For $n$ qubits, we require only $n$ linearly independent stabilizers
to uniquely define a state, an exponential reduction compared to the
description of an arbitrary pure quantum state. The standard cluster
state description uses stabilizer operators (entangling observables)
$S^a$ on a regular lattice of the form,
\begin{equation}
\hat{S}^{a} = X^{a} \bigotimes_{N(a)} Z^{b},
\label{eq:stabilizers}
\end{equation}
where a Pauli $X$ operator acts on qubit $a$ and $Z$ acts on the set
$b$ of neighbouring qubits to $a$, i.e. those sharing an edge with $a$
in the associated graph~\cite{JPhysAMathTheor.43.025301}. We shall be
primarily interested in square or triangular lattices, reflecting the
cluster states easily created in optical lattices. The set of
stabilizer operators describing a cluster state is not unique and we
exploit this in order to generate stabilizer operators with reduced
cross-talk.

\subsection{Fidelity of Cluster States}

Experimentally, determining the closeness of the actual to the desired
state is an important issue. One measure of closeness between the
ideal pure state $|\psi \rangle$ and the actual (mixed) state, $\rho$,
is the fidelity that is defined as
\begin{equation}
\mathcal{F}=\sqrt{\langle \psi | \rho | \psi \rangle}
\label{eq:fidelity}
\end{equation}
Note some author's define the fidelity as
$\mathcal{F}^{'}=\mathcal{F}^{2}$~\cite{Jozsa1994}.

Reconstructing $\rho$ through full quantum state
tomography~\cite{PhysRevA.40.2847} and then calculating $\mathcal{F}$
using Eq. \eqref{eq:fidelity} becomes infeasible with more
than a few qubits due to the exponential number of
parameters. However a lower bound on $\mathcal{F}$ can be achieved by
using certain measurements that are linear in
$n$~\cite{PhysRevA.74.052302}. By defining the following operator
\begin{equation}
\hat{S}_{\mathcal{S}}=\frac{1}{2}\left[\left(\sum_{a=1}^{n} \hat{S}^{a}\right)
-(n-2)\mathbb{1}\right],
\end{equation}
a lower bound of fidelity is given by the expectation of this operator with $\rho$
\begin{equation}
  \mathcal{F}^{2}(\rho , \psi) \geq \langle \hat{S}_{\mathcal{S}} \rangle_{\rho}.
\label{eq:fidelityexp}
\end{equation}
Hence we only require $n$ expectation values of $\{\hat{S}^{a}\}$ in order to estimate the right hand side of Eq. \eqref{eq:fidelityexp}.

\subsection{Measurement and Cross-Talk}

\begin{figure}[h!]
\centering
\subfloat[ ]{
\includegraphics[width=0.35\columnwidth]{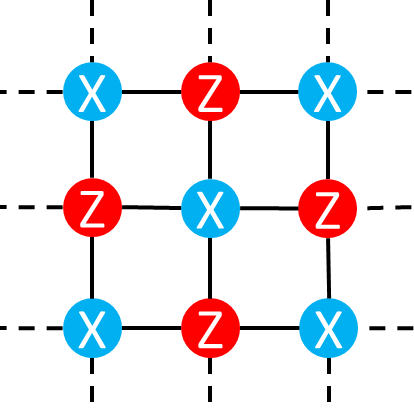}
\label{fig:checkerboard1}
}
\subfloat[ ]{
\includegraphics[width=0.35\columnwidth]{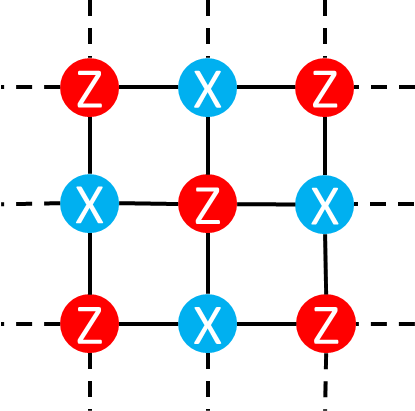}
\label{fig:checkerboard2}
}
\\
\subfloat[ ]{
\includegraphics[width=0.35\columnwidth]{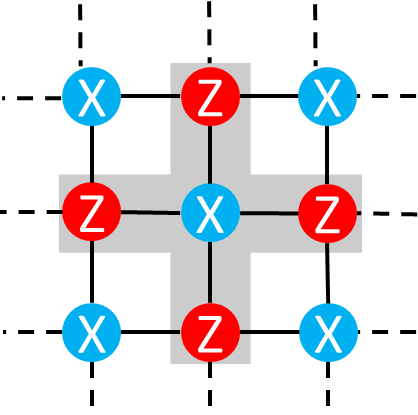}
\label{fig:centralqubit}
}
\\
\subfloat[ ]{
\includegraphics[width=0.3\columnwidth]{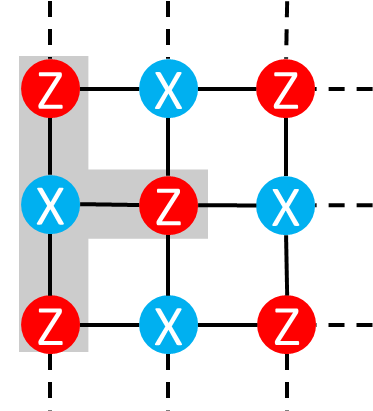}
\label{fig:edgequbit}
}
\subfloat[ ]{
\includegraphics[width=0.3\columnwidth]{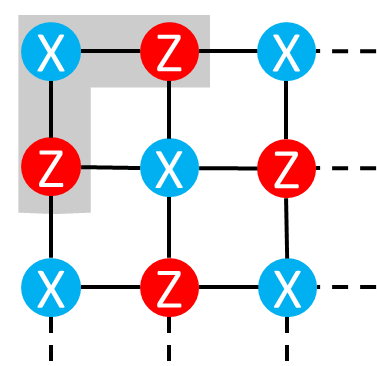}
\label{fig:cornerqubit}
}
\caption{(Colour online) Stabilizer operator Pauli measurement
  patterns. The two patterns of physical measurements performed upon
  the qubits in the lattice are illustrated in
  \protect\subref{fig:checkerboard1} and
  \protect\subref{fig:checkerboard2}. By multiplying the $\pm 1$
  results of these individual operator measurements we can calculate
  the expectation value of the stabilizer operator applied to any
  qubit in the lattice. In \protect\subref{fig:centralqubit},
  \protect\subref{fig:edgequbit} and \protect\subref{fig:cornerqubit}
  the shaded grey areas shows the measurement results that are
  multiplied together to calculate the stabilizer operator on a
  central, edge and corner qubit respectively.}
\label{fig:checkerboard}
\end{figure}

The above method requires that the expectation values of the
stabilizer operators be determined. One could try to measure the
stabilizer operator observables directly but this is an entangled
measurement and difficult to perform in practice. Alternatively, one
can synthesize the measurement value from separate measurements of the
Pauli operator on each qubit and multiplying the results gained during
each instance of the experiment. Averaging over many runs gives an
estimate of the expectation value. To obtain the expectation values
for all the standard stabilizer operators (Eq.~\ref{eq:stabilizers})
for a cluster state on a square lattice, it is sufficient to perform
two measurement patterns as indicated in Fig.\ref{fig:checkerboard}.

Experimentally, such patterns may be problematic. In the case of
optical lattices the natural qubit measurement basis is $Z$,
other measurements directions are produced by unitary
rotations before a $Z$ measurement. These rotations can either be
applied simultaneously to all qubits, or individually with the aid of
an addressing laser beam. However, the beams have waist sizes on the
order of the lattice spacing, hence neighbouring qubits may pick up
unwanted evolutions. This can in principle be ameliorated by the use
of composite pulses~\cite{PhysRevA.67.042308} but this adds
additional complexity and is undesirable. A simple mitigation would be
find measurements that would reduce or eliminate the degree of
crosstalk. This can be achieved by exploiting the non-uniqueness of
stabilizer sets describing a given cluster state.

\section{Stabilizer Operator Sets}

\subsection{Equivalent sets of Stabilizer Operators}

To uniquely define a cluster state we need only specify $n$ linearly
independent stabilizer operators, this choice of $n$ operators is not
unique. We use this fact to our advantage in creating an equivalent
set of stabilizer operators with a reduced amount of cross-talk. Two
sets of stabilizers specify the same state $|\psi \rangle$ if
$\hat{S}^{a}$ is related to $\hat{S}'^{a}$ by a non-singular binary
matrix $m=(m_{jk})$ with $m_{jk} = 0 \text{ or } 1$,
\begin{equation}
\hat{S}'^{j}=\prod_{k=1}^{n} (\hat{S}^{k})^{m_{jk}},
\end{equation}
this construction of the new equivalent set also allows for
reconstruction of the canonical set $\mathcal{S}$ found using
Eq.~\eqref{eq:stabilizers}.

\textbf{Example:} 
\begin{figure}[h]
\includegraphics[width=.3\columnwidth]{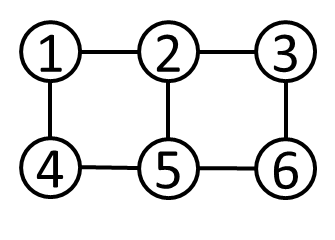}
\caption{$3 \times 3$ qubits cluster state}
\label{fig:32}
\end{figure}

Fig.~\ref{fig:32} shows a $3 \times 2$ cluster state with the following
canonical stabilizer operators, found using Eq.~\eqref{eq:stabilizers}
\begin{equation}
\begin{split}
s_{1} &= X_{1} Z_{2} \mathbb{1}_{3} Z_{4} \mathbb{1}_{5} \mathbb{1}_{6} \\
s_{2} &= Z_{1} X_{2} Z_{3} \mathbb{1}_{4} Z_{5} \mathbb{1}_{6} \\
s_{3} &= \mathbb{1}_{1} Z_{2} X_{3} \mathbb{1}_{4} \mathbb{1}_{5} Z_{6} \\
s_{4} &= Z_{1} \mathbb{1}_{2} \mathbb{1}_{3} X_{4} Z_{5} \mathbb{1}_{6} \\
s_{5} &= \mathbb{1}_{1} Z_{2} \mathbb{1}_{3} Z_{4} X_{5} Z_{6} \\
s_{6} &= \mathbb{1}_{1} \mathbb{1}_{2} Z_{3} \mathbb{1}_{4} Z_{5} X_{6} 
\end{split}
\end{equation}

An equivalent set of stabilizer operators for this cluster state can
be specified by the non-singular matrix $m$
\begin{equation}
m=\begin{pmatrix}
1 & 1 & 0 & 0 & 0 & 0\\
0 & 1 & 0 & 0 & 0 & 0 \\
0 & 0 & 1 & 1 & 0 & 0 \\
0 & 0 & 0 & 1 & 0 & 0 \\
0 & 0 & 0 & 0 & 1 & 1 \\
0 & 0 & 0 & 0 & 0 & 1 
\end{pmatrix}
\end{equation}
which corresponds to the new set
\begin{equation}
\begin{split}
s_{12}&=X_{1}Z_{1} Z_{2}X_{2} Z_{3} Z_{4} Z_{5} \mathbb{1}_{6}\\
s_{2}&=Z_{1} X_{2} Z_{3} \mathbb{1}_{4} Z_{5} \mathbb{1}_{6}\\
s_{34}&=Z_{1} Z_{2} X_{3} X_{4} Z_{5} Z_{6}\\
s_{4}&=Z_{1} \mathbb{1}_{2} \mathbb{1}_{3} X_{4} Z_{5} \mathbb{1}_{6}\\
s_{56}&=\mathbb{1}_{1} Z_{2} Z_{3} Z_{4} X_{5}Z_{5} Z_{6}X_{6}\\
s_{6}&=\mathbb{1}_{1} \mathbb{1}_{2} Z_{3} \mathbb{1}_{4} Z_{5} X_{6} 
\end{split}
\end{equation}
this set still stabilizes the cluster state shown in Fig.~\ref{fig:32}, conversely
the canonical set can be specified from these operators.

\subsection{Construction of cross-talk-free stabilizer operators}

\begin{figure}[h!]
\subfloat[ ]{
\includegraphics[width=0.3\columnwidth]{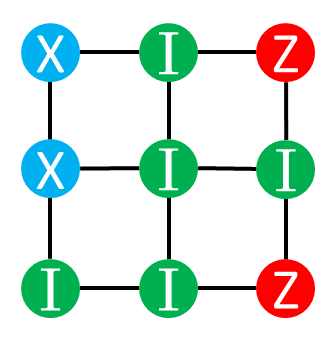}
\label{fig:ctf2}
}
\quad \quad \quad
\subfloat[ ]{
\includegraphics[width=0.3\columnwidth]{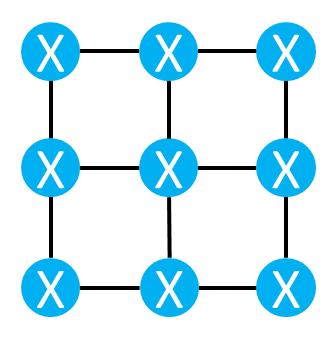}
\label{fig:ctf1}
}
\caption{(Colour online) Examples of ideal cross-talk free stabilizer
  operators. \protect\subref{fig:ctf2} is an example of a CTF
  stabilizer operator still with two types of operator, to physically
  achieve this measurement with no cross-talk we could globally rotate
  all the qubits in the lattice by $\frac{\pi}{2}$ so the $X$
  operators can be measured and locally address the $Z$ operators to
  rotate them back, the locally addressed qubits are far enough away
  from each other that this will not cause any cross-talk
  errors. \protect\subref{fig:ctf1} is an example of a homogeneous CTF
  as is only contains one type of operator this can be achieved simply
  by globally rotating all the qubits in the lattice.}
\label{fig:CTF}
\end{figure}

The cross-talk in the stabilizer operators comes from having to
measure two different operators, namely $X$ and $Z$, on adjacent
qubits. To counter this we have two options: either we split up the
$X$ and $Z$ operators so they are no longer applied to adjacent
qubits, or we find stabilizer operators requiring only one type of
operator. Note that this does not have to be a $Z$ operator as we can
globally rotate all the qubits in the lattice without
cross-talk. There are potentially various kinds of cross-talk free (CTF)
stabilizer operators (Fig.~\ref{fig:CTF}) each with different properties.

We can immediately
  rule out the first idea of separating the $X$ and $Z$ operators by
  examining how the stabilizer operators are constructed from
  Eq.~\eqref{eq:stabilizers}. If we consider a one dimensional line of
  $n$ qubits and we would like to have the $X$ operator as far away as
  possible from the $Z$ operator so the stabilizer operator would look
  like Eq.~\eqref{eq:XZ}
\begin{equation}
X_{1} \mathbb{1}_{2} \mathbb{1}_{3} \dots
\mathbb{1}_{n-2} \mathbb{1}_{n-1} Z_{n},
\label{eq:XZ}
\end{equation}
To create this pattern we follow the steps shown 

\begin{equation}
\begin{split}
\textit{Step 1} \quad &X_{1} Z_{2} \mathbb{1}_{3} \dots \mathbb{1}_{n} \\ \nonumber
\textit{Step 2} \quad &X_{1} \mathbb{1}_{2} X_{3} Z_{4} \mathbb{1}_{5} \dots \mathbb{1}_{n} \\ \nonumber
&\vdots \\ \nonumber
\textit{Step $n_{even}$} \quad &X_{1} \mathbb{1}_{2}X_{3} \dots \mathbb{1}_{n-2} X_{n-1} Z_{n}\\ \nonumber
\textit{Step $n_{odd}$} \quad &X_{1} \mathbb{1} X_{3} \dots \mathbb{1}_{n-1} X_{n} \\ \nonumber
\end{split}
\end{equation}

When $n$ is odd there is no possible way to eliminate the $Z$ operator
on qubit $n$ without performing a trivial operation that just
reassigns the $XZ$ pairing somewhere else. However when $n$ is even we
find no $Z$ operators which leads nicely to our second idea of using
only one type of operator in our stabilizer operators.

As we have just seen it is possible to eliminate all the $Z$ operators
in our stabilizer operator, let us now look to see if we can also
eliminate all the $X$ operators. When we apply
Eq.~\eqref{eq:stabilizers} to each qubit in the system there will only
be one $X$ operator applied to each qubit over the whole set of
operators 
this means that we cannot eliminate the $X$ operators as it
is only possible to multiply the $X$ operator with a $Z$ or an
$\mathbb{1}$ operator. Given that the only way to find CTF stabilizer
operators is with a single type of operator we have to look for those
with only $X$ operators. We will call these homogeneous cross-talk free
(HCTF) stabilizer operators.

Ideally, we would want all the stabilizers that describe a cluster
state to be CTF. As we have shown above we can only generate HCTF
stabilizers. A set of only HCTF stabilizers would only be able to
uniquely specify an eigenstate of $X$ of each qubit. We are thus
forced to include stabilizers that have cross-talk, we will
see how to choose sets that minimize this.

\section{Minimizing Cross-Talk}

To completely define our cluster state using stabilizer operators we
must be able to recreate the canonical set defined by
Eq. \eqref{eq:stabilizers}. It is clear that our set of HCTF
stabilizer operators cannot do this alone as there is no way to create
$Z$ Pauli operators from $X$ and $\mathbb{1}$ operators and so we must
include some non-CTF stabilizer operators. This means there will be
some cross-talk in the system, but if we are intelligent about our
choice of non-CTF stabilizer operators we can see that this can be
brought in at a minimum. Here we define a cross-talk penalty $P_{CT}$
that shows how many pairs of $XZ$ Pauli operators share an edge in any
stabilizer operator. Initially we define the following

\begin{equation}
\begin{split}
P_{CT}^{T}&=\sum P_{CT}(\hat{S}^{a}),\quad S^{a} \in \mathcal{S}^{'} \\
A=\sum_{jk} a_{jk} E_{jk},
\quad F_{X}^a&=\sum_{j} x_{j}^a E_{j},
\quad F_{Z}^a=\sum_{k}z_{k}^a E_{k},\\
F_{XZ}^a&=\sum_{jk}x_{j}^a z_{k}^a E_{jk},
\end{split}
\label{eq:defs}
\end{equation}
where $A$ is the adjacency matrix of the cluster state with $A_{jk}=1$
when qubits $j$ and $k$ share an edge and $A_{jk}=0$ otherwise, and
$E_{jk}$ is a basis matrix $(E_{jk})_{mn}=\delta_{jm}
\delta_{kn}$. For each stabilizer $\hat{S}^a$, the vector $x_{j}^a$
specifies the position ($x_j=1$, otherwise $0$) of $X$ operators,
similarly $z_{j}^a$ for the $Z$ operators. The $E_{j}$ and $E_{k}$ are
both basis vectors with the $jth$ and $kth$ element as $1$ otherwise
$0$, and $F_{XZ}^a$ is the outer product of $F_{X}^a$ and $F_{Z}^a$.

Using the definitions in Eq. \eqref{eq:defs} we define
$B^a$ by taking the Hadamard product ($\circ$) of $A$ with $F_{XZ}^a$
\begin{equation}
B^a=A \circ \left(F_{XZ}^a\right),
\end{equation}
The Hadamard product or entry wise product \cite{hadamardproduct} is
formed by $B_{jk}^a=A_{jk}(F_{XZ}^a)_{jk}$. The cross-talk penalty
$P_{CT}^a$ for a stabilizer is now
\begin{equation}
  P_{CT}(\hat{S}^a) = Tr[(B^a)^{T} B^a]
=\sum_{jk} (x_{k}^s)^{2} (z_{k}^s)^{2}A_{jk}^{2}
\end{equation}
where $A_{jk},x_{k},z_{j}=0,1$.

\begin{figure}[h]
\subfloat[ ]{
\includegraphics[width=.27\columnwidth]{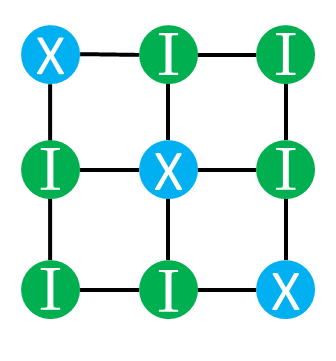}
\label{fig:pctnew1}
}
\subfloat[ ]{
\includegraphics[width=.27\columnwidth]{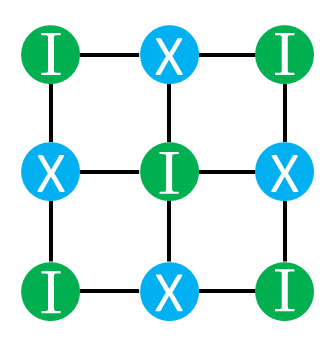}
\label{fig:pctnew2}
}
\subfloat[ ]{
\includegraphics[width=.27\columnwidth]{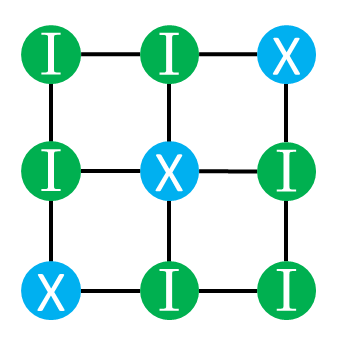}
\label{fig:pctnew3}
}
\\
\subfloat[ ]{
\includegraphics[width=.27\columnwidth]{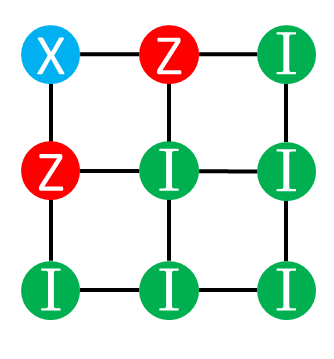}
\label{fig:pctnew4}
}
\subfloat[ ]{
\includegraphics[width=.27\columnwidth]{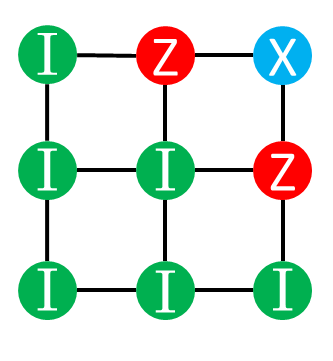}
\label{fig:pctnew5}
}
\subfloat[ ]{
\includegraphics[width=.27\columnwidth]{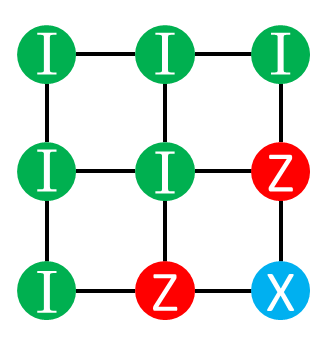}
\label{fig:pctnew6}
}
\\
\subfloat[ ]{
\includegraphics[width=.27\columnwidth]{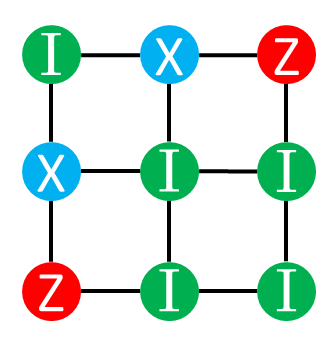}
\label{fig:pctnew7}
}
\subfloat[ ]{
\includegraphics[width=.27\columnwidth]{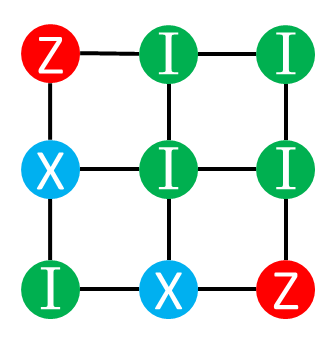}
\label{fig:pctnew8}
}
\subfloat[ ]{
\includegraphics[width=.27\columnwidth]{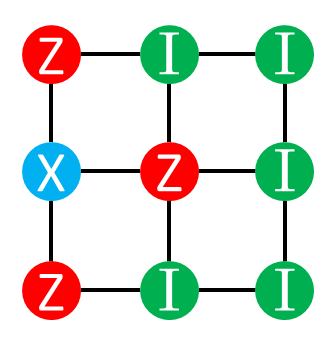}
\label{fig:pctnew9}
}
\caption{The nine stabilizer operators with a reduced $P_{CT}^{T}=13$
  that defines the $3 \times 3$ qubit cluster state.}
\label{fig:pctnew}
\end{figure}

If we take $3 \times 3$ cluster state as an example, the
$P_{CT}^{T_{c}}=24$ if we were to individually measure each of the
canonical stabilizer operators, whereas if we use a set of stabilizer
operators that include the HCTF set and a subset of choice CT
stabilizer operators (Fig.\ref{fig:pctnew}) then the
$P_{CT}^{T_{new}}=13$ which is a big improvement.

\begin{figure}[h]
\subfloat[ ]{
\includegraphics[width=.2\columnwidth]{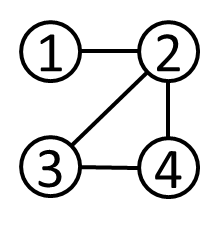}
\label{fig:extendedcrosstalklattice}
}
\subfloat[ ]{
\includegraphics[width=.2\columnwidth]{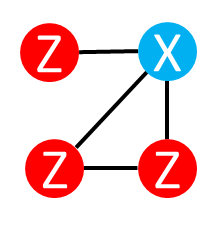}
\label{fig:extendedcrosstalkstabilizer}
}
\subfloat[ ]{
\raisebox{25pt}{
$A= \begin{bmatrix}
0 & 1 & 0 & 0\\
1 & 0 & 0 & 1\\
0 & 0 & 0 & 1\\
0 & 1 & 1 & 0
\end{bmatrix}$
}
\label{fig:adjacency}
}
\caption{(Colour online) A graph state of a modified cluster state. By
  changing the connectivity of the graph state we can see how the
  physical distance between the qubits is important when considering
  the impact of cross-talk. In
  \protect\subref{fig:extendedcrosstalklattice} qubits $2$ and $3$ are
  joined but the distance between them is greater than the range of
  cross-talk, this means when we apply a stabilizer operator to qubit
  2 (in \protect\subref{fig:extendedcrosstalkstabilizer}) the
  cross-talk between qubits $2$ and $3$ is not included, this is
  achieved by modifying the adjacency matrix
  \protect\subref{fig:adjacency}. This model could be further extended
  by allowing the elements of $A$ to have a value between $0$ and $1$
  to get a more accurate value for the impact of the cross-talk, in
  this case we would take the square root of the value between $0$ and
  $1$ as $A_{jk}$.}
\label{fig:extended}
\end{figure}

This model can be extended to incorporate more complicated definitions
of how the cross-talk interferes with the system. For example if qubits
are connected by an edge but the physical distance between them is
greater than the range of the cross-talk it would not be included in
the $P_{CT}$ (Fig.~\ref{fig:extended}).

\section{Cross-Talk-Free Stabilizers}

In arbitrary shaped cluster states it is hard to find HCTF stabilizer
operators, this problem is similar to that of tiling problems which
are non local and NP complete
\cite{RBerger,RMRobinson,10.1109/SFCS.1978.9}. Given these
difficulties for the general problem we have identified, for simple
shapes, patterns and observations starting with the simplest example
of a square lattice. This then leads on to fixed width triangular
lattices that share many similarities with the squares.

Though we cannot create HCTF stabilizer operators for all shapes of
lattice, in certain cases it is possible. In particular we specify how
many HCTF stabilizer operators can be found in general for square,
rectangular and fixed width triangular lattices, and we present
algorithms for generating HCTF stabilizer operators for constant width
lattices.

\subsection{Shapes of lattices that allow HCTF}

\begin{figure}[h]
\subfloat[ ]{
\includegraphics[width=0.25\columnwidth]{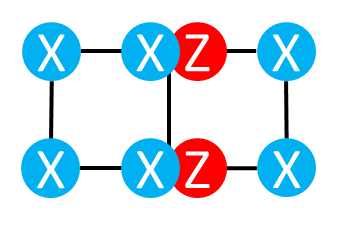}
\label{fig:nontriv1}
}
\subfloat[ ]{
\includegraphics[width=0.3\columnwidth]{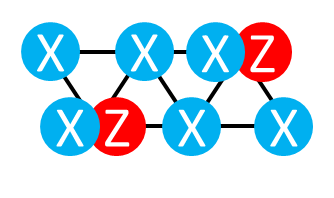}
\label{fig:nontriv2}
}
\caption{(Colour online) Graphical representation that it is not
  possible to construct non-trivial HCTF stabilizers for any
  shaped/connectivity of lattice. \protect\subref{fig:nontriv1} shows
  a square connectivity lattice with extra $Z$ operators that do not
  cancel due to the shape of the overall
  lattice. \protect\subref{fig:nontriv2} shows a triangular
  connectivity lattice again with extra $Z$ operators that do not
  cancel due to the shape of the overall lattice.}
\label{fig:nontriv}
\end{figure}

It is possible to create cluster states in many shapes with different
kinds of connectivity, however not all these shapes and connectivity
of lattices allow for non-trivial HCTF stabilizer operators
(Fig.~\ref{fig:nontriv}). This is due to the number of edges connecting
each of the nodes in the cluster state, a node with an odd number of
edges cannot be surrounded by stabilizer operators as this will lead
to $Z$ Pauli operators that cannot be cancelled.

\subsection{HCTF stabilizer operators in fixed width lattices}

\begin{figure}[h]
\subfloat[ ]{
\includegraphics[width=.22\columnwidth]{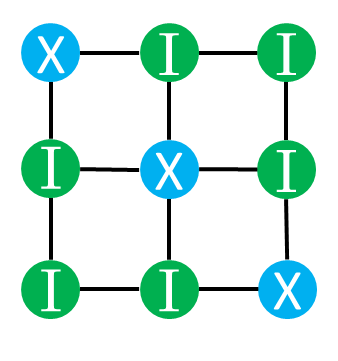}
\label{fig:stab31}
}
\subfloat[ ]{
\includegraphics[width=.22\columnwidth]{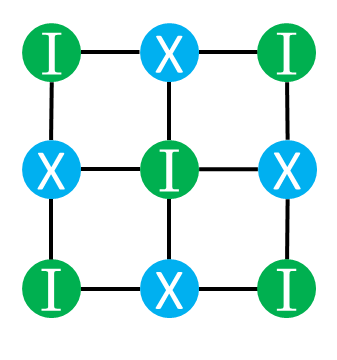}
\label{fig:stab32}
}
\subfloat[ ]{
\includegraphics[width=.22\columnwidth]{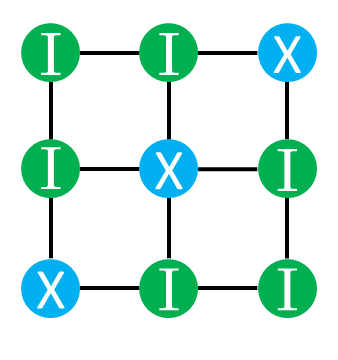}
\label{fig:stab33}
}
\subfloat[ ]{
\includegraphics[width=.22\columnwidth]{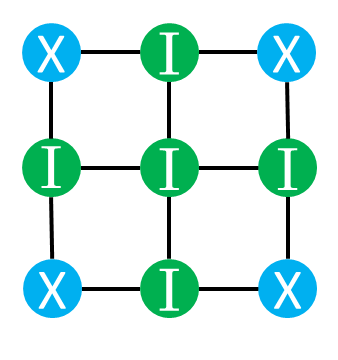}
\label{fig:combinationstab}
}
\caption{(Colour online) The three linearly independent HCTF
  stabilizer operators in a $3 \times 3$ square lattice created by
  applying a single stabilizer operator individually to each of the
  qubits in the initial row are shown in \protect\subref{fig:stab31},
  \protect\subref{fig:stab32} and \protect\subref{fig:stab33}. Where
  as \protect\subref{fig:combinationstab} shows a HCTF stabilizer
  operator that is not in the canonical set that is constructed using
  \protect\subref{fig:stab31} and \protect\subref{fig:stab33}}
\label{fig:stab3}
\end{figure}

When considering square lattices with square connectivity and $n
\times n$ qubits we find there are $n$ linearly independent HCTF
stabilizer operators. This is due to the construction of the HCTF
stabilizer operators, if we approach the lattice row by row and apply
a single stabilizer operator in the first row, there are $n$ possible
places for this stabilizer to start, then by considering the lattice
one row at a time we see that each of these cases leads to an
independent HCTF stabilizer operator (Fig.~\ref{fig:stab3}). It is clear
that they are each linearly independent as they do not share any
qubits in the initial row. These linearly independent operators are
the building blocks for all HCTF stabilizer operators, all other HTCFs
are made up of combinations of these (Fig.~\ref{fig:stab3}).

\begin{figure}[h]
\includegraphics[width=0.95\columnwidth]{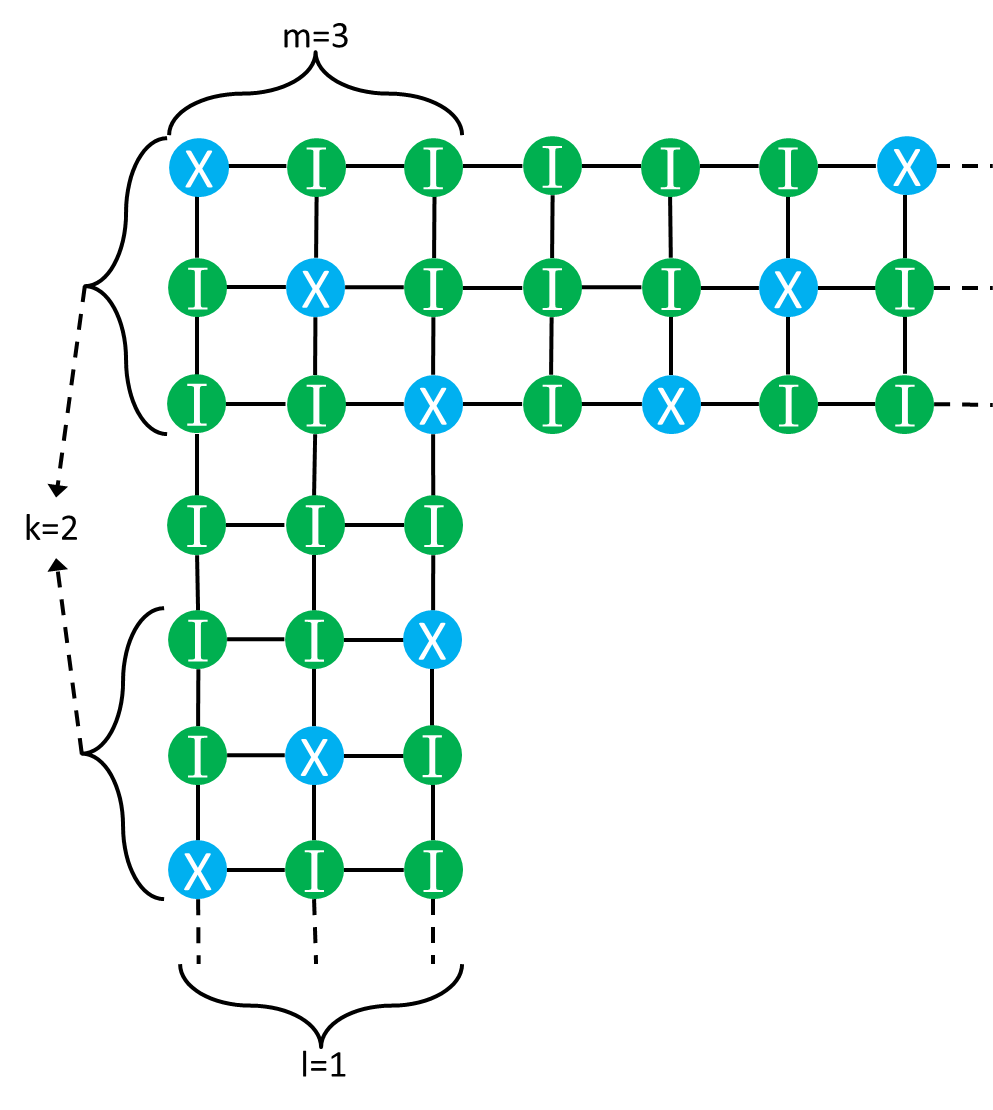}
\caption{(Colour online) Extended lattice for HCTF stabilizer
  operators. It is possible to extend the lattice of a HCTF stabilizer
  operator but this extended lattice must be of the form $(km + (k-1))
  \times (lm + (l-1))$ in order to still create a HCTF stabilizer
  operator. As can be seen from the figure the small individual
  pattern of $m \times m$ is flipped each time it is repeated.}
\label{fig:unsymmetricklm}
\end{figure}

Leading on from the square lattices with square connectivity we find
that it is possible to extend lattices beyond just $n \times n$ qubits
and still create a set of HCTF stabilizer operators, the shape of
these extended lattices is restricted to the form $(km + (k-1)) \times
(lm + (l-1))$ (Fig.~\ref{fig:unsymmetricklm}).

\begin{figure}[h]
\subfloat[ ]{
\includegraphics[width=.3\columnwidth]{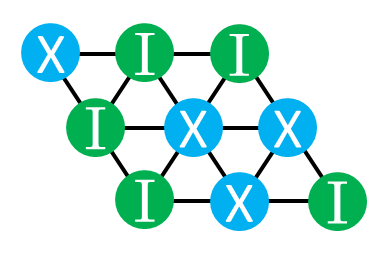}
}
\subfloat[ ]{
\includegraphics[width=.3\columnwidth]{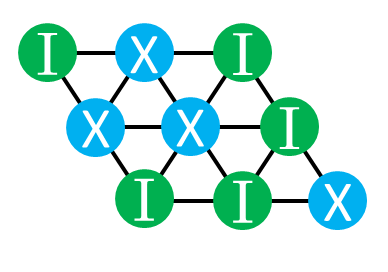}
}
\subfloat[ ]{
\includegraphics[width=.3\columnwidth]{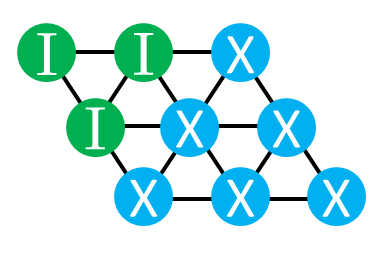}
}
\caption{(Colour online) The canonical generating set of HCTF
  stabilizer operators on a fixed width three qubits triangular
  connectivity lattice. This generating set is formed by applying a
  stabilizer operator individually to each of the qubits in the
  initial row and following algorithm~\ref{alg:algorithm2}.}
\label{fig:trifix3}
\end{figure}

Fixed width triangular connectivity lattices also have no extra
degrees of freedom so we can find a similar deterministic algorithm as
we did for the squares by considering the lattice from an fixed
initial pattern on the first row of the lattice
(algorithm~\ref{alg:algorithm2}). In this case we find $n$ HCTF
stabilizer operators for a lattice of width $n$ qubits
(Fig.~\ref{fig:trifix3}). The HCTFs again appear as squashed square shapes
this allows all the nodes in the graph to have an even number of edges
meaning it is possible for all the $Z$ operators to cancel out.

\subsection{Algorithm for finding HCTF stabilizer operators}

As we have already discussed we can find HCTF stabilizer operators by
starting from a pattern of stabilizer operators in the initial row of
the cluster state, this allows us to create an algorithm to find any
HCTF stabilizer operators where the initial row has been defined.
\begin{algorithm}[h]
 \KwData{initial row $a_{r}= \{0_{0}, 0_{1}, \dots , 1_{m}, 0_{m+1}\}$, 
\\ \qquad dummy row $a_{r-1}= \{0_{0}, \dots , 0_{m+1}\}$}
 \KwResult{HCTF Stabilizer Operator}
 rownumber=3\;
 \While{Number of X operators in the current row $\neq$ 0}{
	\quad  $a_{r+1}^{c}= a_{r}^{c-1} + a_{r}^{c+1} + a_{r-1}^{c} \mod 2$\;
	\quad  for $c=2 \dots m$\;
	\quad  Print $a_{r+1}^{c}$ from $c=2 \dots m$\;
	\quad  Count X operators in the row\;
	\quad $a_{r+1}=\{0, a_{r+1}^{2}, a_{r+1}^{3}, \dots , a_{r+1}^{m}, 0\}$\;
	\quad  rownumber=rownumber+1\;
 }
 \caption{Algorithm to form HCTF stabilizer operators in a rectangular
   lattice given an initial first row. $a_{r}^{c}$ denotes the qubit
   in row $r$, column $c$. The number of qubits in the initial row is
   $m$, we add additional $0$ elements at the start and end of the
   initial row and a dummy row that sits above our initial row to
   ensure the equation holds. The program finds the configuration of
   the $X$ operators in each row of the HCTF stabilizer operator and
   shows how many rows is necessary to complete the HCTF.}
\label{alg:algorithm}
\end{algorithm}

\section{Triangle triangular connectivity lattices}

\begin{figure}
\subfloat[ ]{
\includegraphics[width=.3\columnwidth]{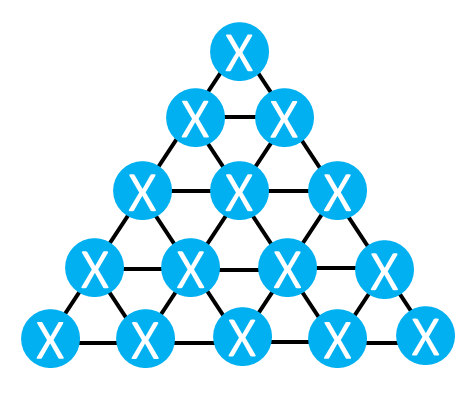}
\label{fig:tristab51}
}
\subfloat[ ]{
\includegraphics[width=.3\columnwidth]{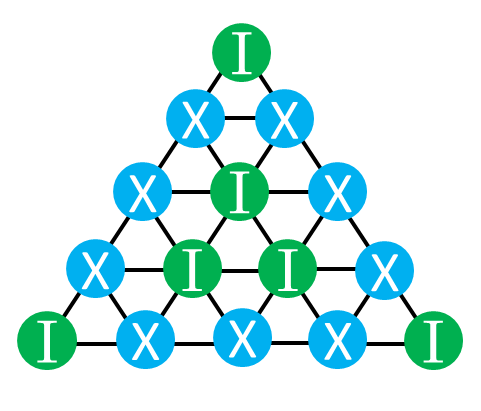}
\label{fig:tristab52}
}
\subfloat[ ]{
\includegraphics[width=.3\columnwidth]{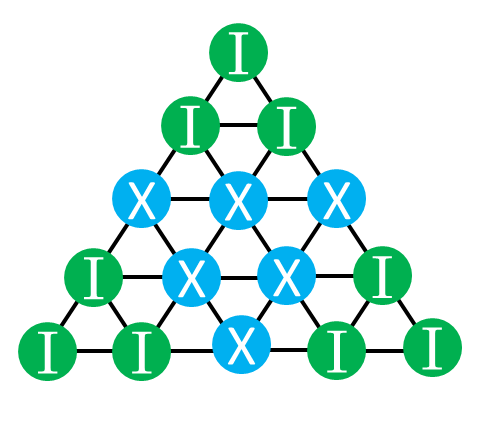}
\label{fig:tristab53}
}
\\
\subfloat[ ]{
\includegraphics[width=.3\columnwidth]{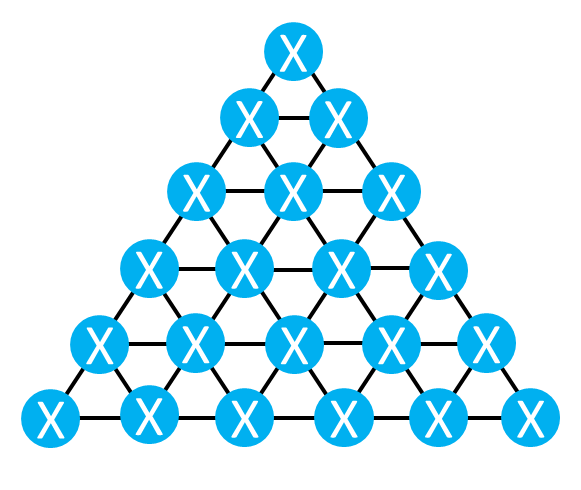}
\label{fig:tristab61}
}
\subfloat[ ]{
\includegraphics[width=.3\columnwidth]{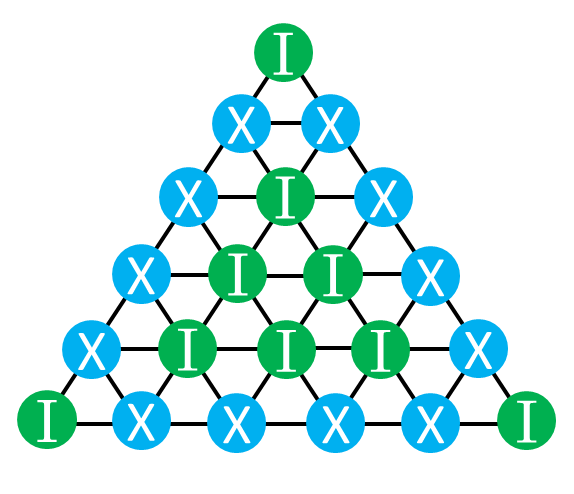}
\label{fig:tristab62}
}
\subfloat[ ]{
\includegraphics[width=.3\columnwidth]{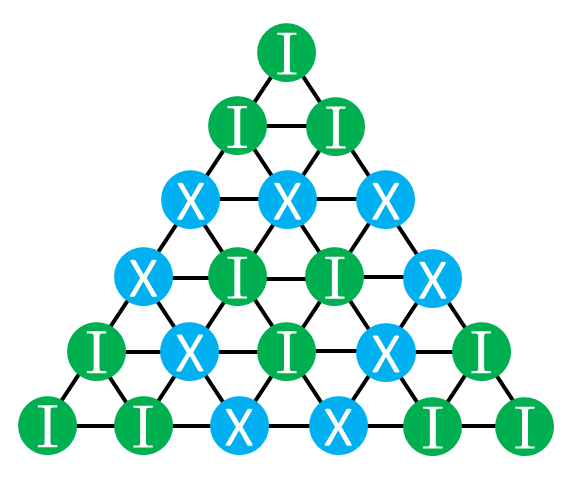}
\label{fig:tristab63}
}
\caption{(Colour online) The canonical set of HCTF stabilizer
  operators. \protect\subref{fig:tristab51},
  \protect\subref{fig:tristab52} and \protect\subref{fig:tristab53}
  show the canonical HCTF stabilizer operators for a $5$ qubit per
  edge triangular lattice. \protect\subref{fig:tristab61},
  \protect\subref{fig:tristab62} and \protect\subref{fig:tristab63}
  show the canonical HCTF stabilizer operators for a $6$ qubit per
  edge triangular lattice. The number of HCTF stabilizer operators for
  a triangle triangular connectivity lattice does not increase when
  going from an odd qubit side length to the next even qubit side
  length. Each HCTF stabilizer operator from an odd sided triangle,
  $r=(2m+1)$ qubits, where $m$ is an integer, has an equivalent HCTF
  stabilizer operator in the next even sided triangle, $r=(2m+1)
  +1$. The symmetry of the pattern does not allow a further HCTF
  stabilizer operator to exist, the centre of the triangle expands
  without affecting the original pattern.}
\label{fig:tristab56}
\end{figure}

Due to the changing degrees of freedom in a triangle triangular
connectivity lattice our previous approach to finding a deterministic
algorithm does not work, and so we considered the lattices
individually using a brute force approach. By looking at every
possible configuration from $1$ qubit to $45$ qubits we find a pattern
in the number of independent HCTF stabilizer operators for each of
these lattice sizes. For a triangle triangular connectivity lattice of
side length $r$ qubits the number of HCTF stabilizer operators is
equal to $\lfloor \frac{r+1}{2} \rfloor$. This $1,1,2,2,3,3\dots$
pattern is interesting as to go from a triangle of odd number of
qubits to even number along each edge does not increase the possible
number of HCTF stabilizer operators (Fig.~\ref{fig:tristab56}).

The canonical set of HCTF can be found by first applying stabilizer
operators to all $r$ qubits along the edges of the lattice, then
completing the pattern internally to eliminate the $Z$ operators, the
second HCTF stabilizer operator is found by applying stabilizer
operators to the qubits along the edges avoiding the corner qubits,
$r=1$ and $r=r$. The next HCTF avoids qubits $r=1,r=2$ and
$r=r,r=r-1$, this pattern is repeated until the last HCTF stabilizer
operator where the stabilizer operator is applied to the central qubit
($r=\frac{r+1}{2}$ for $r$ odd) or qubits ($\frac{r}{2}$ and
$\frac{r+2}{2}$ for $r$ even) (Fig.~\ref{fig:tristab56}).

\begin{algorithm}[h]
 \KwData{initial row $= \{0_{0}, 0_{1}, \dots , 1_{m}, 0_{m+1}\}$, 
\\ \qquad dummy row $= \{0_{0}, \dots , 0_{m+1}\}$}
 \KwResult{HCTF Stabilizer Operator}
 rownumber=3\;
 \While{Number of X operators in the current row $\neq$ 0}{
	\quad  $b_{r+1}^{c}=b_{r+1}^{c-1}+b_{r}^{c-1}+b_{r-1}^{c}+b_{r-1}^{c+1}+b_{r}^{c+1} \mod 2$\;
	\quad  for $c=2 \dots m$\;
	\quad  Print $b_{r+1}^{c}$ from $c=2 \dots m$\;
	\quad  Count X operators in the row\;
	\quad $b_{r+1}=\{0, b_{r+1}^{2}, b_{r+1}^{3}, \dots , b_{r+1}^{m}, 0\}$\;
	\quad  rownumber=rownumber+1\;
 }
 \caption{Algorithm to form HCTF stabilizer operators in a fixed width
   triangular lattice given an initial row. $b_{r}^{c}$ denotes the
   qubit in row $r$, column $c$. The number of qubits in the initial
   row is $m$, we add additional $0$ elements at the start and end of
   the initial row and a dummy row that sits above our initial row to
   ensure the equation holds. The program finds the configuration of
   the $X$ operators in each row of the HCTF stabilizer operator and
   shows how many rows is necessary to complete the HCTF.}
\label{alg:algorithm2}
\end{algorithm}

\section{Conclusion}

\label{sec:conclusion}

By verifying our cluster state using stabilizer operators we come
across problems such as cross-talk in the physical measurement
process. By adapting the measurements we perform on the system we can
reduce these affects to give a more realistic value to our
measurement.

Exploring different shaped lattices we find simple algorithms that
produce sets of linearly independent HCTF stabilizer operators for
lattices where the connectivity and number of qubits remain constant
on each row. Given that it is not possible to find a complete set of
HCTF stabilizer operators to describe our cluster state we consider
how best to choose from the non-CTF stabilizer operators to reduce the
overall affect the cross-talk has on the system, and introduce a
rating system to compare the stabilizer operators.

We could look at different types of lattices such as hexagons
but here we run into the same problems as the triangle shaped
triangular lattices. The connectivity and the number of qubits change
in every row no matter how the hexagonal lattice is constructed, this
changes the degree of freedom each time we add a new row meaning it is
not possible to find a deterministic algorithm.

\begin{figure}[h!]
\subfloat[ ]{
\includegraphics[width=0.3\columnwidth]{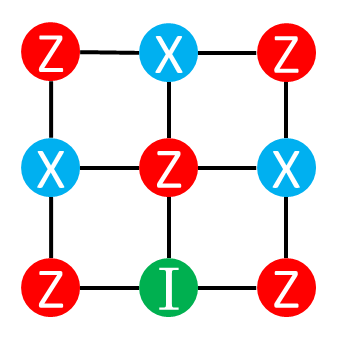}
\label{fig:new2}
}
\subfloat[ ]{
\includegraphics[width=0.3\columnwidth]{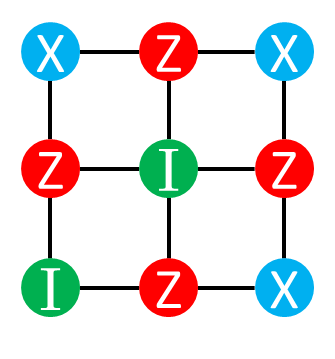}
\label{fig:new3}
}
\subfloat[ ]{
\includegraphics[width=0.3\columnwidth]{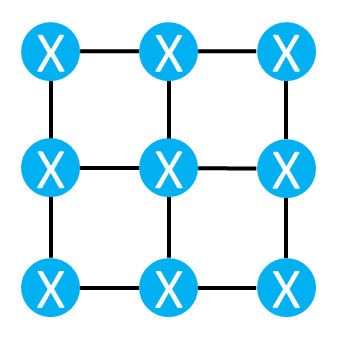}
\label{fig:new1}
}
\caption{(Colour online) Patterns of stabilizer operators that will
  allow a full construction of the canonical set of stabilizer
  operators for a $9$ qubit square lattice. By reducing the cross-talk
  in our measurements, $P_{CT}^{T}=15$, we increase the number of
  patterns that must be measured. The systematic error of the
  measurement is reduced by including the HCTF stabilizer operators
  meaning the result is closer to the value without cross-talk but to
  get a statistical value for the measurement we either take the same
  amount of measurements of each pattern as before therefore
  increasing the time taken to produce the measurement result with the
  same precision, or we reduce the number of repeated measurements
  taken over all the patterns, reducing the accuracy but still
  providing a better measurement.}
\label{fig:new}
\end{figure}

Now we have an algorithm to form HCTF stabilizer operators and a
rating system to determine the best set of modified stabilizer
operators for a rectangular lattice it is important that we also
consider the cost of these improvements, in the initial set up it
takes two different patterns of measurement to construct all the
stabilizer operators of our system, these measurements are heavily
affected by cross-talk reducing the reliability of the result. However
when we introduce our improved set of measurements we find that there
is more than two different patterns of measurements needed to
construct all the stabilizer operators of our system
(Fig.~\ref{fig:new}). As each pattern is measured many times to build up
good statistics, there may be a trade-off between fewer patterns and
more measurements per pattern, or more patterns with reduced crosstalk
but worse statistics.

\begin{figure}[h]
\subfloat[ ]{
\includegraphics[width=.3\columnwidth]{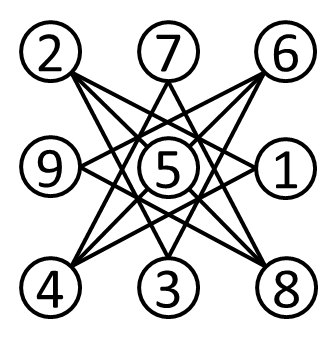}
\label{fig:rearrangedlattice}
}
\subfloat[ ]{
\includegraphics[width=.3\columnwidth]{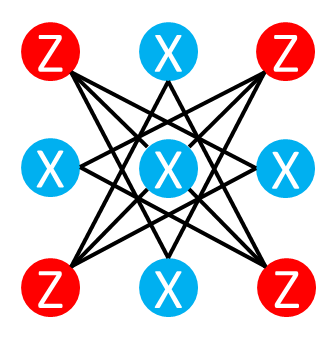}
\label{fig:checkerboardrearrangedone}
}
\subfloat[ ]{
\includegraphics[width=.3\columnwidth]{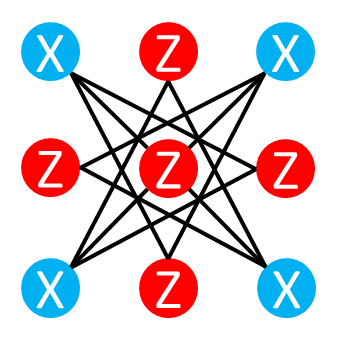}
\label{fig:checkerboardrearrangedtwo}
}
\caption{(Colour online) Rearranged lattice for cluster state
  generation. The qubits in \protect\subref{fig:rearrangedlattice}
  that share an edge have been physically moved so that the distance
  between them is greater than that cross-talk range making these
  newly positioned individual stabilizer operators cross-talk
  free. \protect\subref{fig:checkerboardrearrangedone} and
  \protect\subref{fig:checkerboardrearrangedtwo} show the checkerboard
  pattern required to reconstruct the stabilizer operator expectation
  values, these patterns are obviously not cross-talk free. However
  this is still an improvement on the checkerboard patterns in
  Fig.~\ref{fig:checkerboard} as $P_{CT}^{T}=16$. The entangling
  operations require shifts along various lattice vectors.}
\end{figure}

To get around the problem of cross-talk in a physical sense we could
construct lattices differently so that the atoms that are connected by
an edge are physically far apart so that when the active rotation is
performed by the addressing beam the pairs of $X$ and $Z$ operators in
the stabilizer operator no longer feel the cross-talk
(Fig.~\ref{fig:rearrangedlattice}). There are a couple of things to note
about this idea, firstly it is important to take into consideration
the complexity of the entangling operations as the more complex the
shape the harder it is to create. It is also important to consider how
many measurements will be needed to reconstruct the stabilizer
operator expectation values in the newly shaped lattice.

When different experimental setups it is important to take into
account the possibility of vacancies in our system and incomplete
measurement~\cite{HallOi2014}. In this circumstance it becomes
important to think about which measurement result to assign to the
vacant result, this is dependent upon the number of Pauli operators in
the stabilizer operator. For example, say we have a $3 \times 3$ qubit
square connectivity lattice then then canonical set of HCTF stabilizer
operators look like those in Fig.~\ref{fig:stab3}, and we have assigned
the measurement result $+1$ to the vacancy measurement, if we measure
the HCTF stabilizer operator Fig.~\ref{fig:stab31} on a completely vacant
state our expectation value of the HCTF stabilizer operator will be
$+1$ which is effectively a ``perfect'' state, when in actual fact this
is the opposite. This means that it is better to assign the $-1$
measurement result to a vacancy measurement and attempt to have as
many odd number of Pauli operators in the HCTF stabilizer operators as
possible as in the worst case scenario this will lead to a $-1$
expectation value flagging this fact that the state is flawed.

\section{Acknowledgements}
We acknowledge useful discussion with V\'{a}clav Poto\v{c}ek. KCSH
acknowledges funding from EPSRC.






\bibliography{bibcross}

\end{document}